\documentclass[aps,prb,superscriptaddress,showkeys,longbibliography,twocolumn,nofootinbib]{revtex4-1}

\usepackage[nolist]{acronym}
\usepackage{amsmath,amssymb}
\usepackage{bm}
\usepackage{booktabs}
\usepackage{graphicx}
\usepackage{xcolor}
\usepackage{hyperref}
\usepackage{float}
\usepackage{overpic}
\usepackage{enumitem}
\usepackage{mathptmx}
\usepackage{etoolbox}
\setlist[enumerate,1]{label=(\roman*)}
\setlist[itemize,1]{label=\textbf{--}, leftmargin=1.5em}
\setlist[itemize,2]{label=\textendash, leftmargin=*}

\newcommand{\bsigma}{\bm{\sigma}}

\newcommand{\angstrom}{\text{\AA}}
\newcommand{\Dex}{\Delta_{\mathrm{ex}}}
\newcommand{\dEN}{\delta_{\mathrm{EN}}}
\newcommand{\muB}{\mu_{\mathrm{B}}}

\begin{document}
%\newsavebox{\invisiblebox}
%\sbox{\invisiblebox}{
\begin{acronym}
\newacro{AFM}{antiferromagnetic}
\newacro{DFT}{density-functional-theory}
\newacro{DOS}{density-of-states}
\newacro{FiM}{ferrimagnetic}
\newacro{FM}{ferromagnetic}
\newacro{GBT}{gradient-boosted trees}
\newacro{LOCO}{leave-one-composition-out}
\newacro{MAE}{mean absolute error}
\newacro{ML}{machine-learning}
\newacro{NM}{non-magnetic}
\newacro{PDOS}{projected density-of-states}
\newacro{RF}{random forests}
\newacro{SQS}{special quasirandom structure}
\newacro{SVM}{Support vector machines}
\newacro{TZP}{triple-$\zeta$ polarized}
\newacro{VEC}{valence electron concentration}
\newacro{CI}{confidence interval}
\newacro{CV}{cross-validation}
\end{acronym}
%}
\title{Orbital fingerprinting of magnetism across the Mn--Ni--Ga Heusler ternary}

\author{Felipe Hawthorne}
\affiliation{Department of Physics, Federal University of Parana,
               R. Evaristo F. Ferreira da Costa, 81530-015, Curitiba, Brazil}
\affiliation{Interdisciplinary Center for Science, Technology, and Innovation (CICTI), Federal University of Parana, Av. Cel. Francisco H. dos Santos, 81530-000, Curitiba, Brazil}

\author{Daniela A. Damasceno}
\affiliation{I-X Centre for AI In Science, Imperial College London, White City Campus, 84 Wood Lane, London W12 0BZ, United Kingdom}
\affiliation{Department of Mechatronics and Mechanical Systems Engineering, University of São Paulo, São Paulo 05508-030, Brazil}

\author{Raphael Tromer}
\affiliation{University of Brasília, Institute of Physics, Brasília, Federal District, Brazil}

\author{Ronaldo Rodrigues Pel\'a}
\affiliation{Distributed Algorithm and Supercomputing Department, Zuse Institute Berlin (ZIB), Takustraße 7, 14195 Berlin, Germany}

\author{Cristiano F. Woellner}
\email[]{woellner@ufpr.br}
\affiliation{Department of Physics, Federal University of Parana,
               R. Evaristo F. Ferreira da Costa, 81530-015, Curitiba, Brazil}
\affiliation{Interdisciplinary Center for Science, Technology, and Innovation (CICTI), Federal University of Parana, Av. Cel. Francisco H. dos Santos, 81530-000, Curitiba, Brazil}

\date{\today}

\begin{abstract}
In Mn--Ni--Ga Heusler alloys, the competing magnetic states are separated by differences of a few meV per atom, so the ground state has to be resolved from the electronic structure and cannot be read off the composition. Moving away from the stoichiometric compounds, where the established rules for Heusler magnetism fall short, increases this difficulty. To overcome it, we introduce orbital fingerprints taken from the projected density of states, and use them as the input to machine-learning models for classification of the magnetic ordering, for the magnetic moment amplitude, for the spin polarization at the Fermi level and for interpolation of the phase diagram. The models are trained on a dataset of 370 spin-polarized first-principles calculations of quasirandom structures covering the ternary, which show good agreement with the magnetic ground states and lattice parameters available in the literature. For magnetic ordering, the top-ranked descriptor is the same quantity found by first-principles calculations to distinguish the phases, suggesting that the fingerprints capture the underlying physics.
\end{abstract}

\maketitle

%% ============================================================
\section{Introduction}
\label{sec:intro}
%% ============================================================
 
Manganese-based Heusler alloys are intermetallic compounds with the generic
formula X$_2$YZ (X or Y being Mn) that crystallize in the cubic $L2_1$ structure and
exhibit a diverse suite of magnetofunctional properties tunable by
composition~\cite{Graf2011x,kubler2000}. These include the magnetic shape
memory effect, which yields large, magnetically driven strains used in solid-state actuators~\cite{Ullakko1996,Sozinov2002}, magnetocaloric effects, characterized by field-driven entropy changes for caloric cooling~\cite{planes2009} and half-metallic ferromagnetism, which provides complete spin polarization of conduction electrons for spintronic applications~\cite{deGroot1983,faleev2017}.
In these Heusler alloys, the magnetic ground state depends
sensitively on the interplay between composition, crystal structure, and electronic
structure~\cite{Graf2011x,kubler2000}. This gives rise to rich phase diagrams where
\ac{FM}, \ac{AFM}, \ac{FiM}, and \ac{NM} orderings compete within narrow energy windows.
 
Among the Mn-based Heusler alloys, the Mn--Ni--Ga ternary alloy is one of the most extensively studied systems~\cite{Sokolovskiy2015,Graf2011x}. For instance, the shape-memory alloy Ni$_2$MnGa undergoes a martensitic transformation from a cubic to a tetragonal phase upon changes in temperature or under an applied magnetic field~\cite{Webster1984,
Ullakko1996}. Similarly, the inverse Heusler Mn$_2$NiGa exhibits a strong coupling between structural and magnetic order, with a FiM ground state
driven by Mn moments located on inequivalent sublattices~\cite{liu2006,
singh2014}.

To better understand the magnetic behavior of these systems, it is instructive to consider the microscopic mechanisms governing magnetism in related Heusler compounds. 
\c{S}a\c{s}\i{}o\u{g}lu~\cite{Sasioglu2004,Sasioglu2008} showed that Curie temperatures across the Ni$_2$MnX family are primarily determined by indirect Mn--Ni--Mn exchange mediated by Ni~$3d$ states. 
In a complementary study, Galanakis~\cite{Galanakis2002a,Galanakis2002b}
identified the role of $sp$--$d$ hybridization that yields
half-metallicity and derived the generalized Slater-Pauling rule for full-Heusler compounds, which relates the total spin moment per formula unit to the valence electron count within a rigid-band picture applicable to stoichiometric ordered full-Heusler compounds.
 
At off-stoichiometric, binary, and elemental compositions, the
Slater--Pauling rigid-band picture breaks down~\cite{faleev2017}. In these cases, 
charge-transfer effects driven by electronegativity become significant, and even small compositional variations can alter the magnetic ordering type. As a result, reliable predictions of the magnetic ground state require methods that explicitly account for band-structure details rather than relying on composition alone.

In this context, \ac{ML} approaches applied to Heusler alloys have gained increasing attention. Broadly, these approaches can be classified into two categories.
First, composition-only models (e.g., random forests and gradient-boosted trees) trained on elemental properties achieve good accuracy for lattice constants and magnetic moments in stoichiometric
systems~\cite{Mitra2022,sanvito2017,Jang2023}. However, by construction, they cannot distinguish between different magnetic orderings with identical stoichiometry, nor can they reliably capture properties sensitive to fine electronic-structure details, such as spin polarization at the Fermi level.
Second, system-specific \ac{ML} interatomic potentials that incorporate magnetic degrees of freedom~\cite{Kotykhov2023,Eckhoff2021} provide a more detailed description, but require extensive training data and are therefore not suited for large-scale ternary composition screening.
As a consequence, neither approach can directly access the exchange splittings and hybridization
gaps that first-principles studies have identified as the microscopic
parameters controlling magnetic ordering~\cite{Sasioglu2004,Galanakis2002a}.
This highlights the need for approaches that explicitly retain electronic-structure information while enabling efficient exploration of composition and ordering.
 
Here, we address these limitations by constructing orbital-resolved \ac{PDOS} fingerprints
as \ac{ML} features. These descriptors explicitly encode the electronic-structure characteristics governing magnetism, such as exchange splitting and orbital hybridization, without relying on the atomic spin moments as input.
Hence, one major objective of this work is to assess how well \ac{PDOS}-based fingerprinting can serve as a systematic bridge between first-principles electronic structure and \ac{ML} predictions across the full composition space of a magnetic ternary system, using the Mn--Ni--Ga system as a representative case.

The paper is organized as follows. Section~\ref{sec:methods} outlines the \ac{DFT} calculations, the construction of descriptors, and the \ac{ML} framework.
Section~\ref{sec:results} first validates the \ac{DFT} dataset against known ground states and identifies magnetocaloric candidates. It then turns to the \ac{ML} models, covering classification of the magnetic ordering, regression of the moment amplitude, an \ac{ML}-interpolated ternary phase diagram, and regression of the Fermi-level spin polarization. Section~\ref{sec:conclusion} summarizes the main findings.

%% ============================================================
\section{Methods}
\label{sec:methods}
%% ============================================================
 
\subsection{First-principles calculations}
\label{sec:dft}
\begin{figure}[h]
    \centering
    \includegraphics[width=0.3\textwidth]{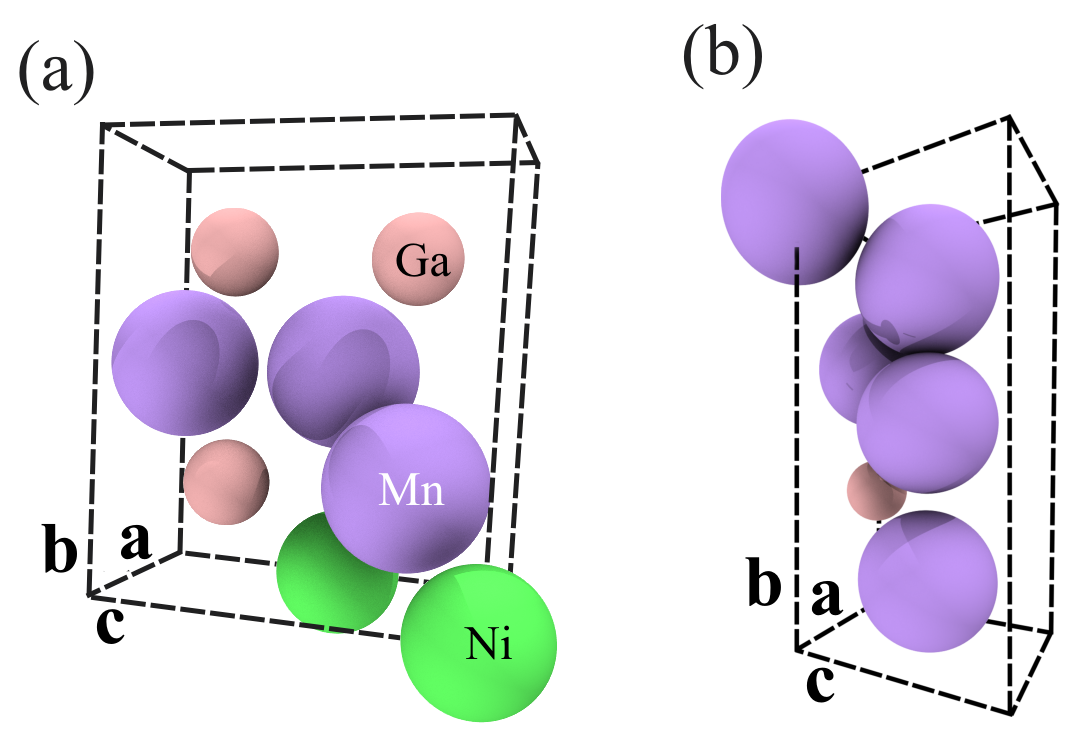}
    \caption{Two representative SQS structures from the dataset:
    (a)~$\text{Mn}_3\text{Ni}_2\text{Ga}_3$ (8 atoms, ternary) and
    (b)~$\text{Mn}_5\text{Ga}$ (6 atoms, Mn-rich binary). Purple,
    green, and pink denote Mn, Ni, and Ga, respectively.}
    \label{fig:ilus}
\end{figure}

A set of 370 structures was generated as cubic \ac{SQS} supercells~\cite{Wei1990,Zunger1990} using the \textsc{icet} package~\cite{Aangqvist2019}, with supercell sizes ranging from 2 to 10 atoms, with 4- and 8-atom cells being the most common.
By construction, \ac{SQS} supercells reproduce the pair and multisite correlation functions of a random alloy. Consequently, none of the generated configurations corresponds to an ordered intermetallic phase, such as the $L2_1$ Heusler or inverse-Heusler structure. The dataset therefore samples compositional disorder across the ternary space rather than a collection of stoichiometric ordered compounds. Figure~\ref{fig:ilus} illustrates two representative structures.

Each structure was independently relaxed starting from both \ac{FM} and \ac{AFM} spin initializations until the maximum force was below $0.05~\text{eV/\angstrom}$, with the lowest-energy configuration being incorporated into the dataset. All calculations were performed within collinear spin-polarized \ac{DFT} as implemented in \textsc{SIESTA}~\cite{soler2002}, using the PBEsol exchange-correlation functional~\cite{Perdew2008}, and norm-conserving Vanderbilt pseudopotentials~\cite{Hamann2013,Garcia2018}. The Kohn--Sham orbitals were expanded in a \ac{TZP} numerical atomic orbital basis~\cite{Artacho1999,Junquera2001}
with a real-space mesh cutoff of 300~Ry. Brillouin-zone integrations employed Monkhorst--Pack~\cite{Monkhorst1976} grids with a $k$-point spacing of $0.5~\angstrom^{-1}$~\cite{jackson_kgrid_2021}.

The converged structures were classified based on the spin
moment per atom, $\{S_{z,i}\}$.
Configurations for which the average local moment
$(1/N)\sum_i|S_{z,i}| < 0.1~\muB$, $N$ being the number of atoms, were assigned a \ac{NM} label.
For the remaining structures, the magnetic ordering was determined through the alignment ratio
\begin{equation}
R_{\text{align}} = |\sum_i S_{z,i}| / \sum_i|S_{z,i}|,
\end{equation}
which distinguishes \ac{FM} ($R_{\text{align}} > 0.5$), \ac{FiM}
($0.1 \leq R_{\text{align}} \leq 0.5$), and \ac{AFM}
($R_{\text{align}} < 0.1$) states. This classification captures the \ac{FiM} behavior prevalent in Mn--Ga-based systems~\cite{Balke2007,Feng2013}. Since $R_{\text{align}}$ varies continuously, the class boundaries are necessarily conventional: fully compensated ferrimagnets ($R_{\text{align}}<0.1$) are assigned to the \ac{AFM} class, and strongly uncompensated ones to the \ac{FM} class.
The resulting dataset spans 90 distinct compositions across the full $\text{Mn}_x\text{Ni}_y\text{Ga}_z$ ternary ($x+y+z=1$),
distributed as 123~FiM (33.2\%), 108~FM (29.2\%), 83~AFM (22.4\%), and 56~NM (15.1\%) configurations. Energies are reported per atom relative to the elemental references through the mixing energy $\Delta E_{\mathrm{mix}} = E - \sum_i x_i E_i^{\mathrm{ref}}$, with $E_i^{\mathrm{ref}}$ the lowest energy from the dataset of a monoatomic structure of species $i$.

All calculations were performed without a Hubbard $U$ correction~\cite{Anisimov1991,Liechtenstein1995}. This choice is justified by the itinerant character of the Mn $3d$ states, whose bandwidth is large enough to place the system in the Stoner regime~\cite{Stoner1938}, where the correlation correction is not the controlling term. Moreover, a single $U$ value cannot consistently describe the wide range of local Mn environments sampled here, from dilute Mn in Ni--Ga to Mn-rich configurations. This choice is further supported by previous studies: both the exchange-interaction analysis of \c{S}a\c{s}\i{}o\u{g}lu~\textit{et~al.}~\cite{Sasioglu2004,Sasioglu2008} and the Slater--Pauling framework of Galanakis~\textit{et~al.}~\cite{Galanakis2002b} employed standard GGA without $U$.

\subsection{Descriptor design}
\label{sec:pdos_extraction}

Four descriptor classes were constructed, namely 33 composition-only features, 64 \ac{PDOS} fingerprints, 20 structural descriptors, and 23 moment-derived descriptors. The first three classes were combined into the 117-feature set used throughout Sec.~\ref{sec:results}, whereas the fourth was used to diagnose target leakage, since these descriptors encode the ordering labels by construction. The remaining seven, taken from the total \ac{DOS} and the Mulliken charges, carry no such information and are retained for the moment regression of Sec.~\ref{sec:mismatch}.

\paragraph{Composition-only}

The composition-only set uses the mole fractions $x_{\mathrm{Mn}}$, $x_{\mathrm{Ni}}$, $x_{\mathrm{Ga}}$, the tabulated Pauling electronegativity $\chi$, metallic radius $r$, atomic mass $m$, $d$-electron count $n_d$ and valence-electron count $Z^{\mathrm{val}}$. Averages are $\bar A = \sum_i x_i A_i$. Associated features calculated using these values were also included in the set, such as the electronegativity spread,
\begin{equation}
  \dEN = \Big[ \textstyle\sum_i x_i (\chi_i - \bar\chi)^2 \Big]^{1/2},
  \label{eq:den}
  \end{equation}
  with $\delta_r$ and $\delta_m$ from $1 - r_i/\bar r$ and $1 - m_i/\bar m$~\cite{YangZhang2012}. The full descriptors are listed in Table~\ref{tab:features}.
\begin{table}[h]
\centering
\caption{Feature description used in the composition-only set.}
\label{tab:features}
\begin{ruledtabular}
\begin{tabular}{@{}p{0.30\linewidth}p{0.55\linewidth}c@{}}
Group & Definition & Quantity \\
\hline
Composition & Mole fractions $x_i$, their squares $x_i^2$,
  and pairwise products $x_i x_j$ & 9 \\
Stoichiometry & Per-cell species counts, total atom count,
  number of distinct elements & 5 \\
Weighted averages & $\bar\chi$, $\bar r$, $\bar m$, $\bar n_d$,
  with $\bar A = \sum_i x_i A_i$ & 4 \\
Valence electron count & $\mathrm{VEC} = \sum_i x_i Z_i^{\mathrm{val}}$
  and its deviations from 6, 7, 7.5, and
  8~\cite{Galanakis2002b,faleev2017,Offernes2008} & 5 \\
Mismatch parameters & Yang--Zhang electronegativity spread $\dEN$
  and normalized radius/mass mismatches $\delta_r$,
  $\delta_m$~\cite{YangZhang2012} & 3 \\
Mixing entropy & $S_{\mathrm{mix}} = -\sum_i x_i \ln x_i$
  and its normalized form & 2 \\
Elemental indicators & Binary flags for the presence of
  Mn, Ni, Ga & 3 \\
Mn-specific ratios & $x_{\mathrm{Mn}} - 0.5$ and
  $x_{\mathrm{Mn}}/(x_{\mathrm{Mn}} + x_{\mathrm{Ni}})$ & 2 \\
\end{tabular}
\end{ruledtabular}
\end{table}

\paragraph{PDOS}

The \ac{PDOS} fingerprints set was built from the orbital- and spin-resolved \ac{DOS} of each system in the dataset.  Four channels were chosen based on their roles in the magnetism of Mn-based alloys~\cite{Sasioglu2004,Sasioglu2008,kubler2000,Galanakis2002a,Galanakis2002b}: ($i$) Mn~$3d$, which carries the dominant local moment and controls the intra-atomic exchange splitting; ($ii$) Ni~$3d$, which mediates the indirect Mn--Ni--Mn exchange responsible for long-range ferromagnetic coupling; ($iii$) Ga~$4p$, which participates in the $sp$--$d$ hybridization that shapes the minority-spin gap in half-metallic compositions; and ($iv$) Mn~$4s$, which probes the extension of the $3d$ exchange field into the itinerant conduction band. Table~\ref{tab:pdos_features} lists all $16$ descriptors for each orbital. 

\begin{table}[h]
\centering
\caption{Orbital-resolved \ac{PDOS} fingerprints, 16 per channel. Definitions are detailed in Appendix~\ref{ap:desc}. Stars${}^*$ indicate calculated for each spin-dependent fingerprints.}
\label{tab:pdos_features}
\begin{ruledtabular}
 \begin{tabular}{@{}p{0.35\linewidth}p{0.5\linewidth}c@{}}
Group & Definition & Quantity \\
\hline
Spectral moments${}^*$ & Spectral weight $I_{X\ell\sigma}$, band center   $\varepsilon_{c,X\ell\sigma}$,  bandwidth $W_{X\ell\sigma}$ and third moment $\gamma_{X\ell\sigma}$& 8 \\
Fermi-level \ac{PDOS}${}^*$ & $g_{X\ell\sigma}(E_F)$, one per spin & 2 \\
Occupied-state counts${}^*$ & $f^{\mathrm{occ}}_{X\ell\sigma}$, the states below
  $E_F$ in the orbital& 2 \\
Occupied-state centers${}^*$ & $\varepsilon^{\mathrm{occ}}_{c,X\ell\sigma}$, the
  band center restricted to occupied states& 2 \\
Spin polarization & $P_{X\ell}(E_F)$ & 1 \\
Exchange splitting & $\Dex^{X\ell}$ & 1 \\
\end{tabular}
\end{ruledtabular}
\end{table}

 The Kohn--Sham eigenstates were projected onto site-centered numerical atomic orbitals~\cite{soler2002,SanchezPortal1995}, and for each chemical species $X$, angular momentum $\ell$ and spin $\sigma$ the contributions of every atom of that species in the cell were summed into a single curve $g_{X\ell\sigma}(\varepsilon)$, calculated from $-20$ to $10$~eV on the absolute Kohn--Sham scale at a spacing of $40$~meV. The sum keeps the fingerprints extensive, so $f^{\mathrm{occ}}_{X\ell\sigma}$ counts the electrons of that species and orbital in the whole cell, while the ratios that define the band center, the bandwidth and the exchange splitting divide the cell size and therefore remain comparable across systems. Moreover, from Table \ref{tab:pdos_features}, we highlight the two features that mostly contribute to the discussion in Sec.~\ref{sec:results}.

First, the $\ell$-resolved exchange splitting
\begin{equation}
\Dex^{X\ell}
= \varepsilon_{c,X\ell\uparrow} - \varepsilon_{c,X\ell\downarrow},
\label{eq:dex}
\end{equation}
defined as the displacement between the majority- and minority-spin band centers of Eq.~(\ref{eq:band_center}), a quantity that measures how far exchange has driven the two spin channels apart in energy. \c{S}a\c{s}\i{}o\u{g}lu~\textit{et~al.}~\cite{Sasioglu2004} showed that the magnetic ordering of Mn-based Heusler alloys is governed by the spin polarization of the conduction electrons and by the position of the unoccupied Mn~$3d$ states relative to $E_F$, making $\Dex$ a physically motivated descriptor for macroscopic ordering. Second, the channel-resolved spin polarization at the Fermi level,
\begin{equation}
P_{X\ell}(E_F)
= \frac{g_{X\ell\uparrow}(E_F) - g_{X\ell\downarrow}(E_F)}
{g_{X\ell\uparrow}(E_F) + g_{X\ell\downarrow}(E_F)},
\label{eq:spin_pol}
\end{equation}
that represents the quantity accessed in spin-polarized photoemission experiments.

\paragraph{Structural class}

For the structural descriptors, geometric and stress values that describe the relaxed cell itself were included. Table \ref{tab:struct_features} lists the 20 descriptors calculated. The Fermi energy was included in this class due to it being a global, per-structure, scalar feature, as well as the reference for the \ac{PDOS} calculations.

\begin{table}[h]
\centering
\caption{Structural descriptors obtained from the relaxed cell.}
\label{tab:struct_features}
\begin{ruledtabular}
\begin{tabular}{@{}p{0.4\linewidth}p{0.45\linewidth}c@{}}

Group & Definition & Quantity \\
\hline
Lattice parameters & Relaxed $a$, $b$, $c$ & 3 \\
Cell angles & $\alpha$, $\beta$, $\gamma$ & 3 \\
Axial ratios & $c/a$, $b/a$ and the tetragonality
  $\max(c/a,\,a/c)$ & 3 \\
Volume & Cell volume $V$ and volume per atom & 2 \\
Residual stress & The stress tensor $\boldsymbol{\sigma}$, and the
  anisotropy $\max(\sigma_{xx},\sigma_{yy},\sigma_{zz})
  - \min(\sigma_{xx},\sigma_{yy},\sigma_{zz})$ & 7 \\
Hydrostatic pressure & $p = -\tfrac{1}{3}\,\mathrm{tr}\,\bsigma$ & 1 \\
Fermi energy & $E_F$ & 1 \\
\end{tabular}   
\end{ruledtabular}
\end{table}

\paragraph{Moment-derived class}

The moment-derived set, with a total of 23 descriptors, was built from the converged Mulliken spin moments and from the total \ac{DOS}, listed in Table~\ref{tab:moment_features}. In addition to aggregate, element-resolved, and dispersion measures of the spin moments, this set includes the spin-resolved total \ac{DOS} at \(E_F\), the corresponding spin polarization \(P^{\mathrm{tot}}(E_F)\), the minimum of the minority-spin \ac{DOS} within \(\pm0.5\)~eV of \(E_F\), and a binary half-metallic flag raised when that minimum falls below \(0.01\)~states/eV. The motivation for this set was to provide an upper bound on the accuracy attainable once a spin-polarized calculation has converged, against which the fraction of that information recovered from the \ac{PDOS} fingerprints can be determined.

\begin{table}[h]
\centering
\caption{Descriptors derived from the converged spin density. \(S_z^{(i)}\) is the Mulliken spin moment of atom \(i\) and \(q_i\) its
Mulliken charge.}
\label{tab:moment_features}
\begin{ruledtabular}
\begin{tabular}{@{}p{0.33\linewidth}p{0.5\linewidth}c@{}}
Group & Definition & Quantity \\
\hline
Aggregate moments & \(\sum_i S_z^{(i)}\), \(\sum_i |S_z^{(i)}|\),
  \(\max_i |S_z^{(i)}|\), and the net and absolute moment per atom & 5 \\
Element-resolved moments & \(\langle S_z\rangle\) and \(\langle |S_z|\rangle\) over each species & 6 \\
Moment dispersion & Standard deviation of \(S_z^{(i)}\) within each
  element & 3 \\
Frustration index & \(f = \sigma\bigl(|S_z^{(i)}|\bigr)\big/
  \bigl\langle |S_z^{(i)}|\bigr\rangle\) & 1 \\
Sublattice imbalance & \(\bigl|\bar{S}_z^{+}+\bar{S}_z^{-}\bigr|\big/
  \max\bigl(|\bar{S}_z^{+}|,|\bar{S}_z^{-}|\bigr)\), with \(\bar{S}_z^{\pm}\)
  the mean moment of the sublattices & 1 \\
Total DOS at \(E_F\) & Spin-resolved total \ac{DOS}, spin polarization, minority-spin minimum, and half-metallic flag & 5 \\
Mulliken charges & \(\max_i |q_i|\) and the standard deviation of \(q_i\)
  & 2 \\
\midrule
\bottomrule
\end{tabular}
\end{ruledtabular}
\end{table}

\subsection{\ac{ML} models}
\label{sec:ml_framework}

Four families were adopted in our benchmarking: 
 \begin{enumerate}
     \item \Ac{GBT}~\cite{friedman2001}: 500 trees, depth 5, learning rate 0.03, subsample 0.8;
     \item \ac{RF}~\cite{breiman2001}: 500 trees, Gini criterion;
     \item \Ac{SVM}~\cite{cortes1995,drucker1997}: RBF kernel, $C = 10$, kernel width
           $1/n_{\mathrm{features}}$, after feature standardization;
     \item $k$-nearest neighbors (KNN)~\cite{cover1967}: $k = 5$, inverse-distance weighting.
 \end{enumerate}

These four classifiers were chosen based on the hypothesis classes each spans. \ac{GBT} and \ac{RF} construct decision boundaries through recursive splits, work directly on mixed-type features and yield importances that rank which descriptors drive the classification. An \ac{SVM} with an RBF kernel finds the boundary that stays as far as possible from the nearest points of each class, after a nonlinear mapping of the features. KNN has no fitted decision function, so each point is classified by the majority label among its nearest neighbors, and its accuracy depends directly on the distance distribution of the feature space. All four were implemented in \textsc{scikit-learn}~\cite{pedregosa2011}, with hyperparameters held fixed across every scenario, so that accuracy reflects the descriptor quality and not tuning.

For the model fitting, all features were standardized to zero mean and unit variance on the
training fold only, with
stratified splitting ensuring each fold reflects the overall class
distribution (33.2\% FiM, 29.2\% FM, 22.4\% AFM, 15.1\% NM). Classification performance was evaluated by stratified 5-fold \ac{CV} with 95\% bootstrap \ac{CI} (200 iterations), and regression performance is reported as $R^2$ and \ac{MAE} from the same 5-fold splitting.

Beyond standard \acp{CV}, we additionally assessed generalization to compositions absent from the training set. To this end, \Ac{LOCO} cross-validation, in which all structures sharing the same composition are held out simultaneously via grouped $k$-fold splitting, was employed. This procedure tests whether the model has learned transferable physics rather than memorized composition-specific correlations, and provides the most rigorous estimate of generalization to unseen compositions.

Finally, feature importances were quantified via permutation importance~\cite{breiman2001}, which is preferred over Gini importance, as the latter is biased toward features with many distinct values~\cite{strobl2007}. We note, however, that permutation importance can become super-additive~\cite{strobl2008} when features are correlated,  which indeed happens to the two leading descriptors of Sec.~\ref{sec:spin_pol}.

   \begin{figure*}[!tp]
    \centering
    \includegraphics[width=0.25\textwidth]{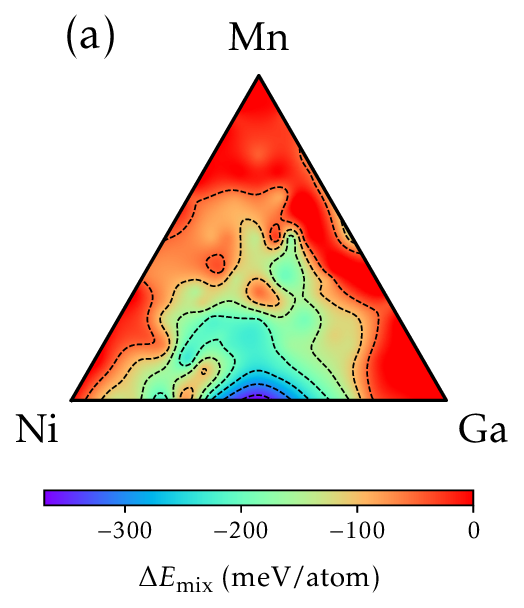}
     \raisebox{1.1cm}{\includegraphics[width=0.25\textwidth]{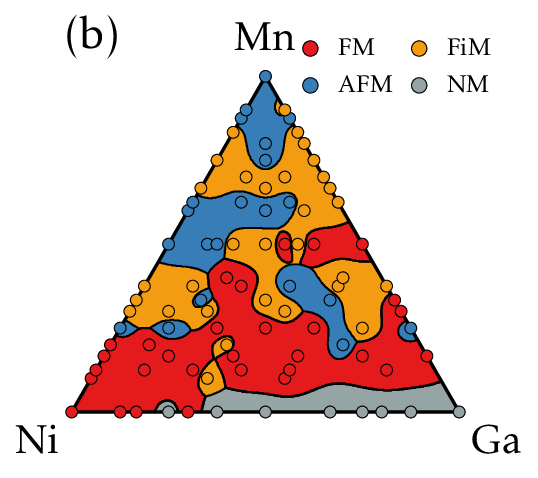}}
    \includegraphics[width=0.25\textwidth]{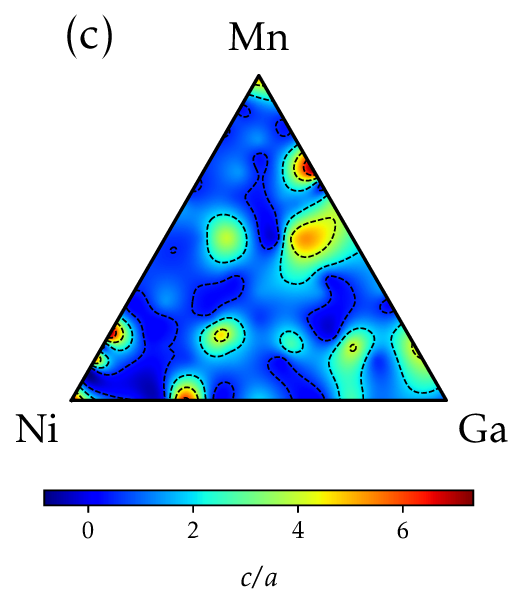}
    \caption{Ternary phase diagrams from \ac{DFT} calculations across the Mn--Ni--Ga composition space: (a)~mixing energy; (b)~magnetic ordering;
    (c)~ relaxed supercell lattice ratio $c/a$. The vertices correspond to elemental phases ($x_A = 1$, with $A = \text{Mn, Ga, Ni}$).}
    \label{fig:ternary_dft}
\end{figure*}

%% ============================================================
\section{Results}
\label{sec:results}
%% ============================================================
\subsection{Outcomes from \ac{DFT} calculations}

\subsubsection{Phase diagrams}

Figure~\ref{fig:ternary_dft} presents ternary phase diagrams derived from \ac{DFT}, showing the compositional dependence of the mixing energy [panel (a)], magnetic ground states [panel (b)], and structural $c/a$ ratios [panel (c)], interpolated from the calculated structures. Along the Ni--Ga binary, the mixing energy [Fig.~\ref{fig:ternary_dft}(a)] reaches a minimum of $-370$~meV/atom at equiatomic NiGa, indicating a pronounced thermodynamic driving force for alloying. The magnetic phase diagram [Fig.~\ref{fig:ternary_dft}(b)] shows that \ac{FM} ordering dominates the Ni-rich corner, \ac{AFM} ordering prevails in the Mn-rich region, \ac{NM} states are most common in the Ga-rich area, and \ac{FiM} ordering spans a broad intermediate region. This distribution reflects the competition between direct Mn--Mn exchange and indirect Mn--Ni--Mn coupling~\cite{Sasioglu2004,Sasioglu2008}.

The $c/a$ ratio [Fig.~\ref{fig:ternary_dft}(c)] is the aspect ratio of the relaxed \ac{SQS} supercell, which the \ac{SQS} construction fixes from the cell size and the site occupation as much as from the composition. While most compositions remain close to $c/a \approx 1$, indicating near-cubic structures, a small number of compositions along the Mn--Ni and Ni--Ga binaries exhibit substantial tetragonal distortions, with $c/a \approx 7$. These regions coincide with compositions where strong magneto-structural coupling and martensitic transformations have been reported experimentally~\cite{Webster1984,Ullakko1996}. For reference, the stoichiometric Heusler Ni$_2$MnGa itself relaxes to $c/a = 1.04$, remaining close to cubic, consistent with its austenite phase above the martensitic transition temperature.

 \subsubsection{Spectral moments}

The spectral moments of Eqs.~(\ref{eq:spectral_weight})--(\ref{eq:third_moment}) were calculated separately in the spin channels from the Mn~$3d$ and Ni~$3d$ orbitals. In this case, the difference between majority- and the minority-spin value of a moment measures the distance between the channels due to exchange. In the case of the band center, this difference is the exchange splitting $\Dex$ of Eq.~(\ref{eq:dex}).

\begin{table}[h]
\caption{Median spin splitting of the Mn~$3d$ spectral moments calculated from the structures containing Mn.}
\label{tab:splitting}
\begin{ruledtabular}
\begin{tabular}{lcccc}
 & \ac{FM} & \ac{FiM} & \ac{AFM} & \ac{NM} \\
\hline
Band center $|\Delta\varepsilon_c|$ [eV] & 1.57 & 0.31 & 0.00 & 0.00 \\
Bandwidth $|\Delta W|$ [eV]              & 0.95 & 0.15 & 0.00 & 0.00 \\
Third moment $|\Delta\gamma|$ [eV$^3$]   & 255  & 49   & 0.3  & 0.1  \\
$N$ structures                            & 95   & 123  & 83   & 29   \\
\end{tabular}
\end{ruledtabular}
\end{table}

Table~\ref{tab:splitting} reports the median splitting of each moment across magnetic orderings, for the spectral weights that carry a spin splitting. The band center, bandwidth, and third moment show a consistent progression, where \ac{FM} states exhibit the largest splitting, \ac{FiM} states approximately one-fifth as much, while \ac{AFM} and \ac{NM} states remain at or near zero. These moments therefore distinguish \ac{FM} from \ac{FiM}, but do not distinguish \ac{AFM} from \ac{NM}. Within the \ac{FM} structures, the Ni~$3d$ band center exhibits a median splitting of 0.28~eV, compared with 1.57~eV for Mn~$3d$, consistent with the smaller induced Ni moments reported for these systems~\cite{Sasioglu2008}.

\subsubsection{Validation against experiment}
\label{sec:validation}

Table~\ref{tab:lit} compares the predicted magnetic ordering at five benchmark compositions against experimental results reported in the literature. These compositions were selected to cover distinct regions of the ternary: Mn-rich, Ni-rich, Ga-rich, ternary interior, and binary.  
All predictions match experiment, providing confidence that the dataset is physically reliable.

Ni$_2$MnGa is correctly identified as \ac{FM} with an energy 23.4~meV/atom lower than the \ac{AFM} configuration. This comparison required selecting the lowest-energy structure across various supercell sizes, with the FM ground state found in a 4-atom cell and the competing \ac{AFM} configuration in an 8-atom cell. Beyond the correct ordering, the same calculation reproduces the experimental structural and magnetic properties, yielding a lattice parameter of 5.76~\AA{} and a magnetic moment of $4.4~\muB$/f.u., in good agreement with the measured values of 5.83~\AA{} and $4.17~\muB$/f.u., respectively~\cite{Webster1984}.

\begin{table}[h]
\caption{Most stable magnetic configurations: comparison between \ac{DFT} (ours) and experimental results. }
\label{tab:lit}
\begin{ruledtabular}
\begin{tabular}{lccc}
Composition & Exp.\ phase & Ground state & DFT  \\
\hline
Mn$_2$NiGa & inv.\ Heusler & FiM~\cite{liu2006} & FiM \\
Ni$_2$MnGa & $L2_1$        & FM~\cite{Webster1984}              & FM   \\
MnNi       & $L1_0$        & AFM~\cite{Kasper1959}              & AFM \\
Mn$_3$Ga   & $D0_{22}$     & FiM~\cite{Balke2007,Feng2013}      & FiM \\
MnGa       & $L1_0$        & FM~\cite{Bither1965,Tanaka1993} & FM \\
\end{tabular}
\end{ruledtabular}
\end{table}

In Ni$_2$MnGa, the \ac{FM} ground state is attributed to indirect Mn--Ni--Mn exchange, which dominates when the Mn sublattice is sufficiently dilute that direct Mn--Mn interactions are rare~\cite{Sasioglu2004}. In contrast, Mn$_2$NiGa converges to a \ac{FiM} state, with no \ac{AFM} solution obtained even from \ac{AFM} initialization, consistent with the strong ferrimagnetic Mn--Mn coupling between inequivalent sublattices characteristic of inverse Heuslers~\cite{liu2006}. MnNi correctly recovers \ac{AFM} ordering, and
Mn$_3$Ga converges to \ac{FiM} with no \ac{FM} ground state, reflecting the tendency of the Mn-rich sublattice toward antiparallel nearest-neighbor
coupling~\cite{Balke2007}.

At the remaining benchmark composition, MnGa, the predicted \ac{FM} ground state agrees with the ferromagnetism of the equiatomic $L1_0$ phase established in bulk~\cite{Bither1965} and in epitaxial films~\cite{Tanaka1993}. We note, however, that bulk samples in this composition range are typically Mn-rich (Mn$_{1+x}$Ga), where excess Mn occupies Ga antisites and couples antiparallel to the Mn sublattice, producing ferrimagnetism~\cite{Zhu2012}. This antisite mechanism requires the ordered $L1_0$ cell and off-stoichiometric occupation of the Ga sites, neither of which is sampled by the cubic \ac{SQS} cells used in this work.

\subsubsection{Magnetocaloric candidates}
\label{sec:mce}

When two magnetic orderings are nearly degenerate in energy, an applied magnetic field can tip the balance between them, producing a large magnetic entropy change $\Delta S_M$. This is the working principle of magnetocaloric refrigeration~\cite{planes2009}. Among the 32 compositions in the dataset where this occurs, four have $\Delta E \leq 5$~meV/atom and eight have $\Delta E < 10$~meV/atom, as shown in Table~\ref{tab:mce}.

\begin{table}[h]
\caption{Magnetocaloric candidates identified from near-degenerate magnetic orderings. $\Delta E$ is the energy difference between the two most favorable magnetic orderings at each composition.}
\label{tab:mce}
\begin{ruledtabular}
\begin{tabular}{lcccc}
Composition & $x_{\mathrm{Mn}}$ & GS & $\Delta E$ [meV/atom] & VEC \\
\hline
\multicolumn{5}{c}{\textit{Sub-5~meV/atom}} \\
MnGa$_3$              & 0.25 & AFM & 0.77 & 4.0 \\
Mn$_9$Ga              & 0.90 & FiM & 0.79 & 6.6 \\
Mn$_3$Ni$_3$Ga$_2$   & 0.38 & FM  & 1.00 & 7.1 \\
Mn$_7$NiGa$_2$        & 0.70 & FiM & 4.07 & 6.5 \\
\hline
\multicolumn{5}{c}{\textit{Sub-10~meV/atom}} \\
Mn (pure)             & 1.00 & AFM & 5.98 & 7.0 \\
Mn$_3$Ni              & 0.75 & FiM & 6.25 & 7.8 \\
MnNi$_3$              & 0.25 & AFM & 6.45 & 9.2 \\
Mn$_7$Ga              & 0.88 & AFM & 6.75 & 6.5 \\
\end{tabular}
\end{ruledtabular}
\end{table}

The four sub-5~meV candidates occupy distinct regions of the ternary and reflect different physical mechanisms underlying the near-degeneracy.
MnGa$_3$ ($\Delta E = 0.77$~meV) lies in the Ga-rich corner, where the dilute Mn concentration weakens direct exchange interactions, leaving competing orderings nearly degenerate in energy. 
At the Mn-rich edge, the $3d$ shell approaches the half-filled $d^5$ configuration (maximizing the moment by Hund's rule) and similarly reduces the energy differences between competing states in Mn$_9$Ga ($\Delta E = 0.79$~meV). Mn$_3$Ni$_3$Ga$_2$ ($\Delta E = 1.00$~meV) represents a ternary composition near Mn$_2$NiGa~\cite{liu2006,singh2014} with $\mathrm{VEC} = 7.1$, placing it within the \ac{FM}/\ac{FiM} competition regime, where direct and indirect exchange interactions are comparable in magnitude. Finally, the fourth candidate, Mn$_7$NiGa$_2$ ($\Delta E = 4.07$~meV), exhibits a somewhat larger (yet still small) energy gap and resides in a similar Mn-rich region dominated by \ac{FiM} ordering.

Among the sub-10~meV candidates, pure Mn stands out: it converged to all four orderings across different supercell geometries, indicating a particularly flat magnetic energy landscape. Furthermore, the $\Delta E$ values for MnGa$_3$ and Mn$_9$Ga fall below the 1--2~meV/atom numerical precision of the \ac{DFT} calculations (Sec.~\ref{sec:dft}), so their exact energetic ordering cannot be resolved without tighter convergence criteria. Precisely this feature, however, makes them strong candidates for the magnetocaloric effect: energy differences this small suggest that an applied magnetic field could switch between the competing orderings. Thermodynamic stability is a separate requirement, and by the mixing energy of Fig.~\ref{fig:ternary_dft}(a) two of the eight candidates lie above the elemental references, MnGa$_3$ at $+7.6$~meV/atom and Mn$_3$Ni at $+0.1$~meV/atom, so those two would need to be stabilized by temperature or by a competing ordered phase before the near-degeneracy could be used. 

\subsection{Outcomes from ML models}

\subsubsection{Classification of magnetic orderings}\label{sec:classification_magnetic_orderings}

This section addresses the classification of magnetic orderings (\ac{FM}, \ac{FiM}, \ac{AFM}, or \ac{NM}) of a given structure using the features introduced in Section~\ref{sec:pdos_extraction}. In this context, four scenarios were designed to assess the individual contribution of each descriptor class, differing in which feature groups enter the model and how generalization is tested.

First, the composition-only scenario includes only mole fractions and tabulated elemental properties, without \ac{DFT}-derived information. Second, the \ac{PDOS} fingerprint scenario augments this baseline with the structural descriptors and the orbital-resolved \ac{PDOS} fingerprints, totaling 117 features; it is named after the \ac{PDOS} fingerprints, as these are the descriptors added relative to the composition-only baseline. Third, the all-features scenario further adds the 23 magnetic-moment-derived descriptors, reaching 140 features. Since these encode the target property by construction, this scenario serves as a leakage diagnostic rather than a genuine predictive model. Finally, the \ac{LOCO} scenario retrains the \ac{PDOS} fingerprint model on compositions held out entirely from training, testing generalization to unseen stoichiometries.

\begin{table}[h]
\caption{Predictive accuracy for magnetic ordering. For each scenario, the best-performing model (\ac{GBT}, \ac{RF}, SVM, KNN) is reported. Brackets indicate 95\% \acp{CI} from bootstrap resampling of the 5-fold \ac{CV}. For the \ac{LOCO} scenario, the mean and standard deviation across composition folds are reported
instead.}
\label{tab:clf}
\begin{ruledtabular}
\begin{tabular}{lccc}
Scenario & Features & Model & Accuracy \\
\hline
Composition-only      & 33  & RF  & 68.1\% [64.3, 71.4] \\
\ac{PDOS} fingerprint & 117 & GBT & 86.8\% [83.8, 89.7] \\
All-features          & 140 & GBT & 93.0\% [91.6, 94.3] \\
\ac{LOCO}             & 117 & GBT & $76.0 \pm 20.1$\% \\
\end{tabular}
\end{ruledtabular}
\end{table}

Table~\ref{tab:clf} reports the best-performing \ac{ML} model for each scenario. \ac{RF} achieves the highest accuracy in the composition-only case, whereas \ac{GBT} outperforms it in the remaining three scenarios. This reflects the complementary strengths of the two algorithms: \ac{RF} handles the moderate-dimensional composition space (33 features) effectively, while \ac{GBT} benefits from its sequential boosting strategy as the feature space expands to 117 or more and the decision boundaries become increasingly complex. The two remaining families gain far less from the fingerprints, as the \ac{SVM} moves from 65.7\% to 68.1\% and KNN from 66.8\% to 71.6\% on going from 33 to 117 features, against 66.8\% to 86.8\% for \ac{GBT}, which indicates that the information the fingerprints add is carried in interactions between descriptors and not in a smoother decision boundary.
 
In the composition-only scenario, the model reaches 68.1\%, with the number of Mn atoms per cell, $n_{\mathrm{Mn}}$, as the most influential feature (permutation importance of 0.053). This trend is already visible in the phase diagram of Fig.~\ref{fig:ternary_dft}(b), where \ac{FM} ordering is concentrated in the Ni-rich corner, giving way to \ac{FiM} and \ac{AFM} orderings as the Mn concentration increases, suggesting that Mn content is the primary compositional variable separating \ac{FM} from non-\ac{FM} states. Adding \ac{PDOS} fingerprints raises the accuracy to 86.8\%, a gain of 18.7 percentage points. 

%The non-overlapping 95\% confidence intervals ([64.3, 71.4]\% versus [83.8, 89.7]\%) indicate that this improvement is not a statistical fluctuation.

The all-features scenario reaches 93.0\% and serves both as an upper bound and as a target-leakage diagnostic. When atomic spin moments are included as features, they dominate the classification model. For instance, $|S_z|$ (importance 0.257), net moment per atom (0.158), and sublattice moment statistics (0.106 and 0.101) absorb virtually all predictive weight, reducing every \ac{PDOS} feature to zero importance. This is the expected signature of target leakage, since the magnetic moments directly define the classification label and therefore contain the answer by construction. The 6.2 percentage point gap between \ac{PDOS} and all-features scenarios quantifies how much information \ac{PDOS} fingerprints recover indirectly through the electronic structure relative to what the moments encode directly.

Withholding entire stoichiometries, the LOCO folds give a mean accuracy of $76.0 \pm 20.1\%$, higher than that from the composition-only set, indicating that much of the \ac{PDOS}--ordering relationship survives the removal of entire stoichiometries from the training set. The spread across folds arises from the uneven distribution of structures across compositions. In particular, the 51 elemental-Mn structures constitute a single composition group and are therefore held out together. Consequently, some folds remove an entire vertex of the composition diagram from the training set, whereas others remove compositions whose neighboring compositions remain represented.

\begin{figure}[h]
    \centering
    \includegraphics[width=0.35\textwidth]{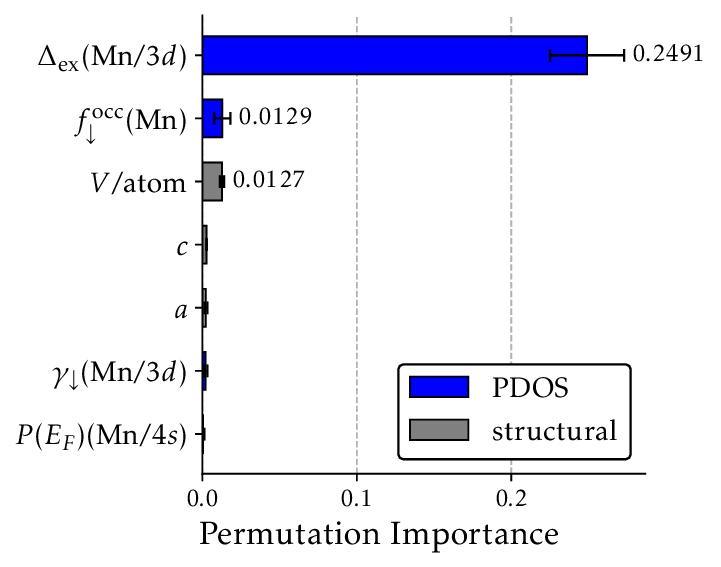}
    \caption{Permutation importance of the top seven features for magnetic
    ordering classification (\ac{GBT}, \ac{PDOS} fingerprint scenario). The
    Mn~$3d$ exchange splitting emerges as the dominant descriptor, with no composition feature
    appearing among the top seven. Bars are colored by descriptor category.}
    \label{fig:clf_importance}
\end{figure}

Figure \ref{fig:clf_importance} depicts the permutation importance for the \ac{PDOS} fingerprints scenario for the GBT model.
A single feature accounts for nearly all of the gain over the
composition-only baseline. The Mn~$3d$ exchange splitting $\Dex$ (importance 0.249) is by far the most informative descriptor. In the itinerant Stoner picture~\cite{Stoner1938}, ferromagnetism emerges when the exchange-energy gain from spin splitting exceeds the kinetic-energy cost of redistributing electrons between spin channels. $\Dex$ is therefore expected to track whether the system sits on the magnetic threshold. In the localized Heisenberg picture~\cite{kubler2000}, the same quantity sets the magnitude of the intersite exchange integrals $J_{ij}$~\cite{Sasioglu2004} that govern both the ordering temperature and the ordering type. The second most relevant feature, Mn~$3d$ minority-spin occupied-state filling $f_\downarrow^{\mathrm{occ}}$ (importance 0.013), further distinguishes among the ordered phases, as \ac{FM} exhibits a partially filled minority band, \ac{AFM} fills both channels equally as the moments cancel across sublattices, and \ac{FiM} falls between these two extremes. 

\begin{figure}[h]
    \centering
    \includegraphics[width=0.4\textwidth]{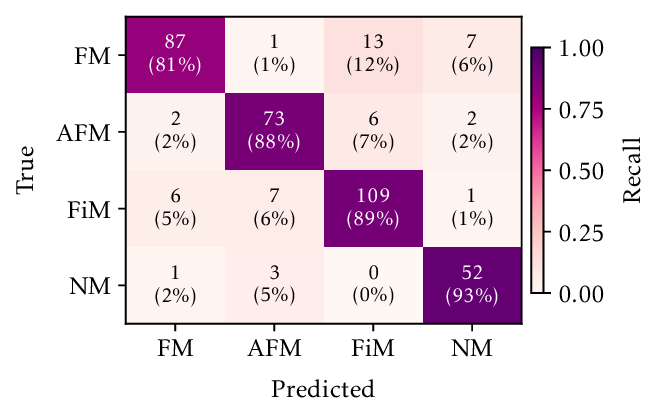}
    \caption{Confusion matrix for the \ac{GBT} classifier in the PDOS-fingerprint scenario (86.8\% accuracy). Rows indicate true labels, columns indicate predictions.}
    \label{fig:confusion_sm}
\end{figure}

Figure ~\ref{fig:confusion_sm} depicts the confusion matrix for the classifier. Among structures whose \ac{DFT} ground state corresponds to a given ordering, the classifier correctly assigns 93\% of \ac{NM}, 89\% of \ac{FiM}, 88\% of \ac{AFM}, and 81\% of \ac{FM} structures to their respective classes. Thus, all four orderings are recovered from the fingerprint with accuracy substantially higher than the 25\% expected from random assignment. 

\ac{NM} marks the absence of local moments altogether and is recovered best, whereas \ac{FM} and \ac{FiM}, which differ only in how completely the sublattice moments compensate, are recovered worst. Confusion between \ac{FM} and \ac{FiM} accounts for 19 of the 49 misclassified structures, including 13 \ac{FM} structures predicted as \ac{FiM}, out of 21 \ac{FM} errors in total, which is what the class definition leads one to expect, since the two are separated by a threshold on the alignment ratio $R_{\rm align}$, a quantity that varies continuously across the dataset, so they overlap in descriptor space wherever the compensation is partial.

 \subsubsection{Magnetic moments and electronegativity spread}
\label{sec:mismatch}

Beyond the classification of the magnetic ordering, the same descriptor set can be used to quantify the moment amplitude. For this task, a leakage-free subset of the all-features scenario of Sec.~\ref{sec:classification_magnetic_orderings} was used, obtained by removing the sixteen descriptors built from the spin moments and keeping the seven based on the total \ac{DOS} and Mulliken charges. With this feature set, the \ac{GBT} model predicts the magnetic moment per atom with $R^2 = 0.898$ and \ac{MAE} of 0.10 $\muB$, against $R^2 = 0.728$ and \ac{MAE} of 0.18 $\muB$ for the composition-only set, suggesting that the \ac{PDOS} fingerprint encodes information about the moment amplitude that the former does not.

\begin{figure}[h]
    \centering
    \includegraphics[width=0.35\textwidth]{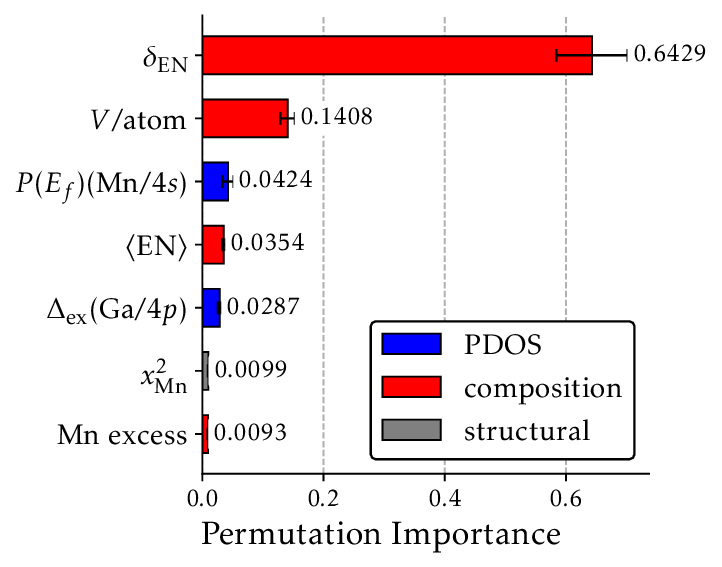}
    \caption{Permutation importance for $|M|$/atom regression
    (leakage-free, \ac{GBT}). $\dEN$ dominates at 0.64, nearly five times
    the next feature (volume per atom, 0.14). \ac{VEC}-based descriptors
    are absent from the top ranks. Bars are colored by descriptor
    category.}
    \label{fig:moment_importance}
\end{figure}

Figure~\ref{fig:moment_importance} depicts the permutation importance for the moment regression, whose dominant feature is not the one anticipated from the Slater--Pauling formalism. The Pauling electronegativity spread $\dEN$ accounts for an importance of $0.643$, nearly five times that of the next most influential feature, volume per atom ($0.141$). The traditional descriptors based on \ac{VEC} and Mn mole fraction~\cite{slater1936,pauling1938,Galanakis2002b}, which highlight the Slater--Pauling rules, do not appear among the top seven. The summed importance of all explicit Mn-related composition features is 0.023, more than 27 times smaller than $\dEN$ alone, indicating that $\dEN$ captures interspecies charge-transfer information that Mn content cannot encode by itself.

\begin{figure}[h]
    \centering
    \includegraphics[width=0.3\textwidth]{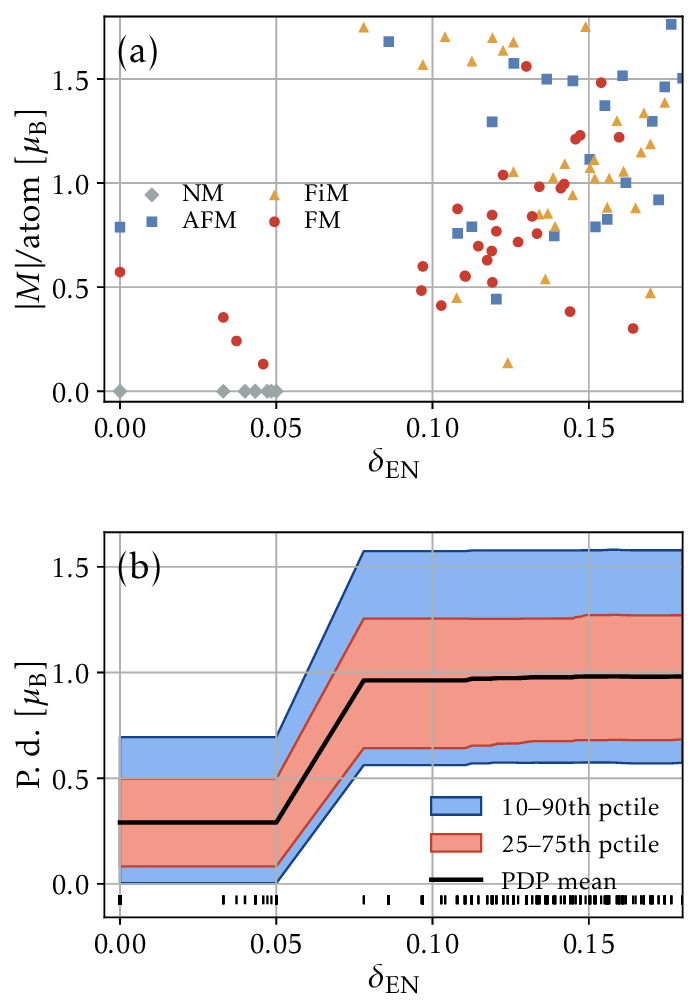}
    \caption{(a)~ground state $|M|$/atom versus $\dEN$ for the 90
    unique compositions, colored by ordering.
    (b)~Partial dependence plot of the \ac{GBT} model on $\dEN$.
    The black curve is the mean predicted $|M|$/atom. Shaded bands
    show the 25th--75th and 10th--90th percentile ranges of
    individual conditional expectation curves.}
    \label{fig:slater_pauling}
\end{figure}

Figure~\ref{fig:slater_pauling}(a) depicts the ground state $|M|$/atom against $\dEN$ for the 90 unique compositions.  The scatter plot supports the positive correlation between $\dEN$ and $|M|$/atom with a Pearson correlation of $r = 0.57$: \ac{NM} states cluster at $\dEN < 0.05$, while the largest moments ($|M| > 1.5~\muB$) appear only above $\dEN \approx 0.1$. \ac{VEC} shows essentially no linear association with $|M|$ over the same set ($r = 0.12$), and the Slater--Pauling prediction $|M| = |\mathrm{VEC} - 6|$~\cite{Galanakis2002b} yields
$R^2 = -4.19$ across the 370 structures, worse than a constant predictor fixed at the dataset mean.
This failure is
expected, as the Slater--Pauling rule was derived for stoichiometric full-Heusler compounds under rigid-band assumptions~\cite{Galanakis2002b} that break down for the off-stoichiometric, binary, and elemental compositions included here. 
 
Figure~\ref{fig:slater_pauling}(b) presents the partial dependence plot~\cite{friedman2001} of the \ac{GBT} model on $\dEN$, revealing a sharp transition learned near $\dEN \approx 0.05$ that corresponds to the \ac{NM}/magnetic boundary visible in the scatter data of panel~(a). Below this threshold, the predicted moment is approximately $0.3~\muB$; above it, the prediction rises to approximately $1.0~\muB$ and saturates. The spread of the individual-conditional-expectation curves above $\dEN \approx 0.05$ spans approximately $1~\muB$ from the 10th to the 90th percentile, indicating that once a composition is magnetic, the moment amplitude is further modulated by secondary features, consistent with the non-negligible roles of volume per atom (importance~$0.141$) and Mn~$4s$ spin polarization ($0.042$) in Fig.~\ref{fig:moment_importance}.

One reading of the dominance of $\dEN$ runs through the local Mn~$3d$ electron population. Ga ($\chi = 1.81$) and Ni ($\chi = 1.91$) are more electronegative than Mn ($\chi = 1.55$) and therefore draw electron density away from Mn, partially depleting its $4s$/$4p$ states and altering the $3d$
filling.  Since the local Mn moment is largest when the $3d$ shell is near half-filling ($d^5$), where the intra-atomic exchange energy is maximized~\cite{Hund1925,kubler2000}, this charge redistribution can modulate the moment magnitude. A larger $\dEN$ drives stronger interspecies charge transfer, pushing the Mn~$3d$ occupation further from the symmetric, nonmagnetic configuration and producing larger moments, in line with the dominant importance value of $0.643$.  The second most important feature, volume per atom, reflects the magnetovolume effect~\cite{kubler2000}, in which lattice expansion narrows the $3d$ bandwidth and increases the Stoner product $I_{\mathrm{xc}} \cdot g(E_F)$, where $I_{\mathrm{xc}}$ is the intra-atomic exchange integral and $g(E_F)$ the total \ac{DOS} at the Fermi level~\cite{Stoner1938}, favoring larger moments through a mechanism independent of electronegativity.
 
\subsubsection{ML-interpolated phase diagram}

\begin{figure}[h]
    \centering
    \includegraphics[width=0.5\columnwidth]{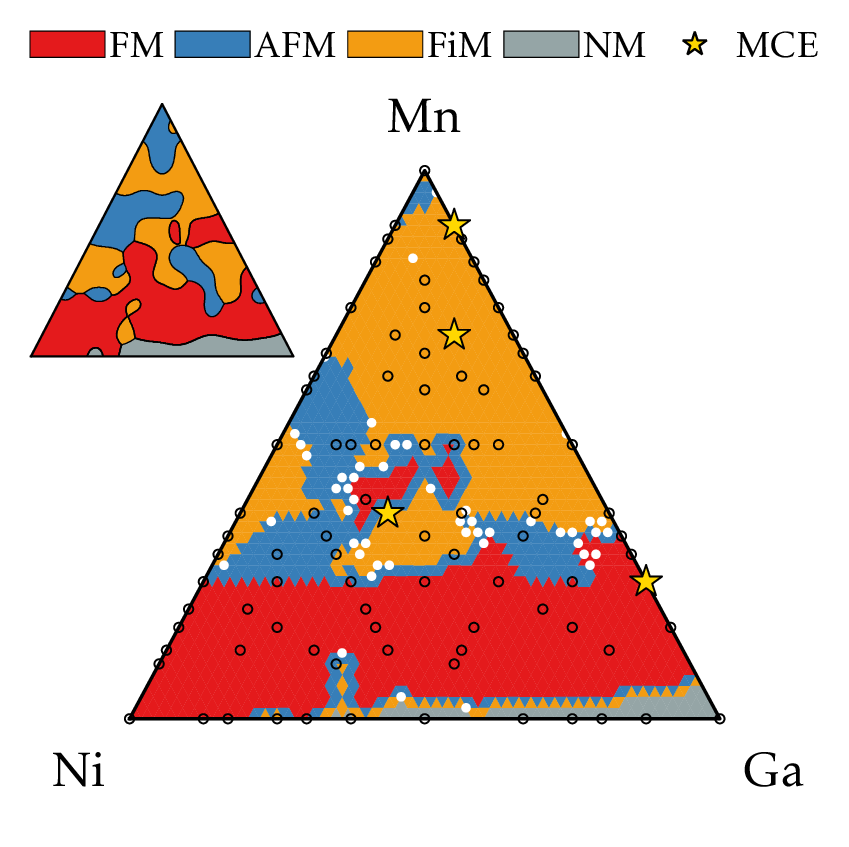}
    \caption{\ac{ML}-interpolated magnetic phase diagram of the Mn--Ni--Ga
    ternary. Grid points are colored according to the magnetic ordering. White dots
    represent low-confidence predictions, while open circles
    indicate compositions computed explicitly with \ac{DFT}. Stars mark the primary magnetocaloric candidates listed in
    Table~\ref{tab:mce}. The phase diagram of Fig.~\ref{fig:ternary_dft}(b) is included as an inset for visual aid.}
    \label{fig:ternary_interp}
\end{figure}

To map the magnetic landscape across the full ternary, a \ac{GBT} classifier with the hyperparameters of Sec.~\ref{sec:ml_framework} was retrained on all 370 structures, using the composition-only descriptor set, and subsequently evaluated on a dense grid of 1326 hypothetical compositions (Fig.~\ref{fig:ternary_interp}). 
The composition-only descriptors are required here, since \ac{PDOS} fingerprints are unavailable for compositions not explicitly computed with \ac{DFT}, so the map inherits the 68.1\% accuracy of that scenario in Table~\ref{tab:clf} . The predicted distribution comprises \ac{FM} (44.0\%), \ac{FiM} (37.6\%), \ac{AFM} (12.5\%), and \ac{NM} (5.9\%). Predictions with maximum class probability below 0.5 (3.8\% of grid points) concentrate at the \ac{AFM}--\ac{FiM} boundary in the Mn-rich sector, consistent with the ground state competition analysis of Sec.~\ref{sec:mce}, which identifies this region as hosting the most closely competing orderings.

Since the predictions in Fig.~\ref{fig:ternary_interp} are based solely on composition, no distinction can be made between structures within the same stoichiometry: the map therefore indicates the most probable ground state at each point of the ternary. Still, the qualitative agreement between this result and Fig.~\ref{fig:ternary_dft}(b) suggests that the ground state ordering is governed more by composition than by the specific atomic configuration. 
Nevertheless, the \ac{FiM} region occupies a larger area of the ternary. This can be attributed to a class imbalance in the training set, where \ac{FiM} along with \ac{FM} orderings account for 231 of the 370 structures, potentially biasing the classifiers.

\subsubsection{Predicting the spin polarization at the Fermi level}
\label{sec:spin_pol}

The final regression target is the total Fermi-level spin polarization, $P^{\mathrm{tot}}(E_F)$,  defined as in Eq.~(\ref{eq:spin_pol}) with the total \ac{DOS} of each spin channel in place of the projected one. The composition-only feature set fails at this task ($R^2 \approx 0$ for \ac{GBT}, with no model family exceeding $R^2 = 0.17$). This failure is expected: the polarization is determined by the spin-resolved \ac{DOS} at the Fermi level and by the possible opening of a minority-spin gap, neither of which is a simple function of the mole fractions. In contrast, with the \ac{PDOS} fingerprint descriptor set, the \ac{GBT} model reaches $R^2 = 0.937$~[0.928, 0.947] with an \ac{MAE} of 0.04, transforming the target from essentially unpredictable to quantitatively accessible.

\begin{figure}[tp]
    \centering
    \includegraphics[width=0.4\textwidth]{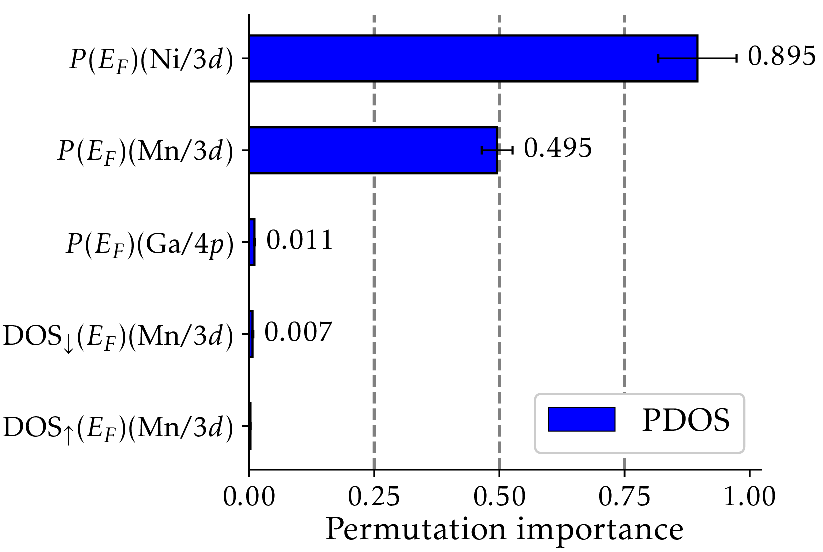}
    \caption{Top-5 permutation-importance
    features for the $P^{\mathrm{tot}}(E_F)$ regression.}
    \label{fig:spinperm}
\end{figure}

The permutation importances of the fitted model are depicted in Fig.~\ref{fig:spinperm}. The two dominant features are the channel-resolved spin polarizations at $E_F$ of the Ni~$3d$ and Mn~$3d$ states, with importances of $0.895$ and $0.495$, respectively, followed distantly by Ga~$4p$ ($0.011$). 

This hierarchy indicates that the model reconstructs the total spin polarization from the $d$-channel contributions, thus making use of the descriptors that mostly contribute to the spectral weight at the Fermi level~\cite{kubler2000, Galanakis2002a}. Part of that reconstruction is close to arithmetic, given that the four channels carry most of the total \ac{DOS} at $E_F$ in this system and a direct sum of them reproduces $P^{\mathrm{tot}}(E_F)$ at $r = 0.995$ with no fitting. The prediction was therefore repeated with every Fermi-level quantity withheld from the feature set, leaving only band centers, bandwidths and occupied-state counts, which still gives $R^2 = 0.521$ against the $R^2 \leq 0.17$ of the composition-only set. The shape of the bands away from $E_F$ therefore carries information about the polarization at $E_F$ that the composition does not.

We note that the two leading importances sum to more than unity ($0.895 + 0.495 = 1.39$). This is the super-additive artifact anticipated in Sec.~\ref{sec:ml_framework}: permutation importance double-counts correlated features~\cite{strobl2007}, and the strong Mn--Ni $3d$--$3d$ hybridization in this system~\cite{Sasioglu2008} correlates the two channel polarizations by construction. 

\section{Conclusions}
\label{sec:conclusion}

This work introduced orbital-resolved \ac{PDOS} fingerprints as descriptors for first-principles-informed prediction of magnetic properties in Heusler alloys. Magnetic ordering is the clearest example, seeing as the top-ranked descriptor is the Mn~$3d$ exchange splitting, which the first-principles analysis identifies as the quantity separating the phases. The classifier reaches 86.8\%, compared with 68.1\% from composition alone. The moment amplitude is predicted with $R^2 = 0.898$, versus $0.728$ from composition and $-4.19$ from the Slater--Pauling rule. For Fermi-level spin polarization, the fingerprints reach $R^2 = 0.937$, compared with $R^2 = 0.17$ for composition-only models. Even without Fermi-level descriptors, $R^2 = 0.521$ is retained, showing that the band shape away from $E_F$ contains information about the polarization at $E_F$.

While composition-only descriptors can suffice for magnetic moments~\cite{Mitra2022,sanvito2017}, they fall short for two tasks central to magnetic materials design: magnetic ordering classification and predicting spin polarization at the Fermi level. 
In this context, the \ac{PDOS} fingerprinting scheme bridges the gap between two extremes. Composition-only screening is fast, but unable to capture ordering- or band-structure-dependent properties. Full electronic-structure calculations are accurate, but computationally expensive. The present approach requires only a single self-consistent DFT calculation per structure, from which all 117 descriptors are extracted with no post-processing beyond the PDOS.
Furthermore, the scheme is system-independent by construction. Extensions to other Heusler ternaries, or to tetragonal distortions, require only the redefinition of the appropriate orbital channels. 

Two limitations of the present approach should be noted. The \ac{SQS} construction samples disordered occupations of a cubic parent lattice, so the ordered site arrangements of the tetragonal phases, $L1_0$ MnGa and $D0_{22}$ Mn$_3$Ga, are absent from the dataset, and with them the antisite ferrimagnetism of Mn-rich Mn$_{1+x}$Ga that develops in those phases~\cite{Zhu2012}. Second, the calculations are collinear, and the supercells hold 2 to 10 atoms, which leaves non-collinear spin textures and any ordering with a longer period out of reach.

In the interest of reproducibility, all DFT input/output files and the extracted PDOS fingerprint dataset are openly available~\cite{github_repo}.

\begin{acknowledgments}
The authors would like to thank CNPq and CAPES for financial support, the Center for Computing in Engineering and Sciences at Unicamp, and the HPC Cluster Coaraci, made available under FAPESP grants 2013/08293-7 and 2019/17874-0, respectively, the Brazilian LNCC (Laboratório Nacional de Computação Científica), for access to the SDumont Cluster, under the SINAPAD/2014 project 01.14.192.00. The authors also gratefully acknowledge the computing time made available to them on the high-performance computer ``Lise'' at the NHR Center NHR@ZIB. This center is jointly supported by the Federal Ministry of Research, Technology, and Space and the state governments participating in the NHR (\url{www.nhr-verein.de}). D.A.D. acknowledges support through an Eric and Wendy Schmidt AI in Science Global Faculty Fellowship.

%\textcolor{red}{dani, tromer}
\end{acknowledgments}

\section*{Data Availability Statement}

The data that support the findings of this study are openly available in
Zenodo~\cite{github_repo}, and are mirrored on GitHub at \url{https://github.com/felipehgc/MnNiGa-dataset}.

\bibliography{refs}
\appendix

\section{Explicit form of the \ac{PDOS} fingerprints}
\label{ap:desc}

The scalars of Table~\ref{tab:pdos_features} are integrals of the summed,
orbital- and spin-resolved density of states $g_{X\ell\sigma}(\varepsilon)$
of species $X$, angular momentum $\ell$ and spin $\sigma$. Except for the two
occupation integrals, they are evaluated by trapezoidal quadrature over the
full tabulated range $[\varepsilon_i,\varepsilon_f] = [-20,10]$~eV of the
absolute Kohn--Sham scale, which contains the entire valence manifold of
every structure in the dataset.

The spectral weight,
\begin{equation}
I_{X\ell\sigma}
= \int_{\varepsilon_i}^{\varepsilon_f}
  g_{X\ell\sigma}(\varepsilon)\, d\varepsilon,
\label{eq:spectral_weight}
\end{equation}
is the total number of states carried by the channel, and it normalizes the
three moments that follow. The band center,
\begin{equation}
\varepsilon_{c,X\ell\sigma}
= \frac{1}{I_{X\ell\sigma}}
  \int_{\varepsilon_i}^{\varepsilon_f}
  g_{X\ell\sigma}(\varepsilon)\, \varepsilon \, d\varepsilon,
\label{eq:band_center}
\end{equation}
locates the mean energy of that spectral weight and enters the exchange
splitting of Eq.~(\ref{eq:dex}). The bandwidth,
\begin{equation}
W_{X\ell\sigma}
= \left[
    \frac{1}{I_{X\ell\sigma}}
    \int_{\varepsilon_i}^{\varepsilon_f}
    g_{X\ell\sigma}(\varepsilon)\,
    (\varepsilon - \varepsilon_{c,X\ell\sigma})^2 \, d\varepsilon
  \right]^{1/2},
\label{eq:bandwidth}
\end{equation}
quantifies the dispersion about $\varepsilon_{c,X\ell\sigma}$, and the third
spectral moment,
\begin{equation}
\gamma_{X\ell\sigma}
= \frac{1}{I_{X\ell\sigma}}
  \int_{\varepsilon_i}^{\varepsilon_f}
  g_{X\ell\sigma}(\varepsilon)\,
  (\varepsilon - \varepsilon_{c,X\ell\sigma})^3 \, d\varepsilon,
\label{eq:third_moment}
\end{equation}
captures its asymmetry.

Two further integrals are taken over the occupied states alone. The
occupation,
\begin{equation}
f_{X\ell\sigma}^{\mathrm{occ}}
= \int_{\varepsilon_i}^{E_F}
  g_{X\ell\sigma}(\varepsilon)\, d\varepsilon,
\label{eq:occ_filling}
\end{equation}
counts the electrons that species $X$ contributes to the cell through the
$\ell$ orbitals of spin $\sigma$, so that the spin difference
$f_{X\ell\uparrow}^{\mathrm{occ}} - f_{X\ell\downarrow}^{\mathrm{occ}}$ gives
the $\ell$-resolved contribution to the magnetic moment of that
species~\cite{Mulliken1955,soler2002}. The occupied-state band center,
\begin{equation}
\varepsilon_{c,X\ell\sigma}^{\mathrm{occ}}
= \frac{1}{f_{X\ell\sigma}^{\mathrm{occ}}}
  \int_{\varepsilon_i}^{E_F}
  g_{X\ell\sigma}(\varepsilon)\, \varepsilon \, d\varepsilon,
\label{eq:occ_center}
\end{equation}
is the analog of Eq.~(\ref{eq:band_center}) restricted to states below the
Fermi level~\cite{Pettifor1995}. Whenever a channel carries no weight, which happens when the corresponding element is absent from the cell, all six quantities are set to zero.

\end{document}